\documentclass[preprint,twocolumn,10pt,a4paper,byrevtex,aip,superscriptaddress,showpacs]{revtex4-1}
\usepackage{graphicx}
\usepackage{amsfonts}
\usepackage{amsmath}
\usepackage{amssymb}
\usepackage{color}

\begin{document}

\title{Transient lateral photovoltaic effect in patterned
metal-oxide-semiconductor films}
\author{J.P. Cascales}
\affiliation{Dpto. Fisica de la Materia Condensada C-III, Instituto Nicolas Cabrera (INC) and Condensed Matter
Physics Institute (IFIMAC), Universidad Autonoma de Madrid, Madrid 28049, Spain}
\author{I. Mart\'inez}
\affiliation{Dpto. Fisica de la Materia Condensada C-III, Instituto Nicolas Cabrera (INC) and Condensed Matter
Physics Institute (IFIMAC), Universidad Autonoma de Madrid, Madrid 28049, Spain}
\author{D. D\'iaz}
\affiliation{Dpto. Fisica de la Materia Condensada C-III, Instituto Nicolas Cabrera (INC) and Condensed Matter
Physics Institute (IFIMAC), Universidad Autonoma de Madrid, Madrid 28049, Spain}

\author{J.A. Rodrigo}
\affiliation{Universidad Complutense de Madrid, Facultad de Ciencias F\'{\i}sicas, Ciudad
Universitaria s/n, Madrid 28040, Spain}
\author{F. G. Aliev*}
\affiliation{Dpto. Fisica de la Materia Condensada C-III, Instituto Nicolas Cabrera (INC) and Condensed Matter
Physics Institute (IFIMAC), Universidad Autonoma de Madrid, Madrid 28049, Spain}
\date{\today }

\begin{abstract}
The time dependent transient lateral photovoltaic effect has been studied
with $\mu s$ time resolution and with chopping frequencies in the kHz range,
in lithographically patterned 21 nm thick, 5, 10 and 20 $\mu m$ wide and
1500 $\mu m$ long Co lines grown over naturally passivated p-type Si (100).
We have observed a nearly linear dependence of the transitorial response
with the laser spot position. A transitorial response with a sign change in
the laser-off stage has been corroborated by numerical simulations. A
qualitative explanation suggests a modification of the drift-diffusion model
by including the influence of a local inductance. Our findings indicate that
the microstructuring of position sensitive detectors could improve their
space-time resolution.

% (*)farkhad.aliev@uam.es
\end{abstract}

\maketitle

When photo-carriers are generated locally in a non-uniformly illuminated
surface or interface, a photovoltaic response can be measured parallel to
the Schottky barrier between two distant lateral contacts. This behavior is
known as the lateral photovoltaic effect (LPE) \cite%
{WALLMARK1957,Lucovsky1960,Boeringer1994,Zhao2006}. The LPE originates from
the diffusion of carriers out of the illuminated area and it has been widely
used to develop high precision position-sensitive detectors (PSD) \cite%
{Yu2010,Jin2013}.

During the last decades, the main way of optimizing the sensitivity of LPE
based PSDs has been achieved by the use of metal-semiconductor junctions,
and quite recently metal-oxide-semiconductor junctions (MOS) with different
types of metals (Ti, Co, \ldots ) \cite{Willens1986,Xiao2008}. The
particular interest shown in Co/SiO$_{2}$/Si structures is related to the
possibility of developing broadband PSD for the visible, ultraviolet or
infrared range \cite{Xiao2007,Kong2008,Qi2010} by adjusting the Co
thickness. Previous studies of MOS structures, including Co/SiO$_{2}$/Si,
investigated the LPE by illuminating a relatively large (about $10\times 10$ 
$mm^{2}$) rectangular samples. A laser beam of a few mW was focused into a
10-50 $\mu m$ spot and the open circuit LPE was measured between two
contacts which extended along opposite edges \cite{Xiao2007,Qi2010,Yu2010}.
In wide devices, the dynamic response is dominated by the barrier
capacitance which results in a unipolar (charge-discharge type) response 
\cite{Kronic1999} which diminishes in amplitude when the chopper frequency
approaches the kHz range \cite{Kabra2008}. Moreover, the steady state LPE
value diminishes substantially when the Co thickness approaches 20 nm \cite%
{Xiao2008}. Recent advances in electron beam lithography have permitted the
development of MOS structures where the LPE can investigated along
patterned, micron wide, metallic line structures. The time dependent
photovoltaic response along such structures might be different from the one
observed in the wide LPE devices.

Our work investigates the lateral photovoltaic effect in lithographically
patterned (21 nm thick, 5, 10 and 20 $\mu m$ wide and 1500 $\mu m$ long) Co
lines. The structures were deposited on a naturally passivated (about 2 nm
SiO$_{2}$) Silicon (100) substrate. More details on preparation and
characterization of samples may be found in \cite{Brems2007,Herranz2009}.

We have studied the transient photovoltaic effect (T-LPE) \cite%
{Dhariwal1976,Vieira1997,Zhao2006} as a response to turning the laser beam
illumination on (steady state, referred to as the ON state) followed by
switching the laser off (decaying regime, referred to as the OFF state) as a
function of the spot position, pulse frequency and power. We observe
peak-like transitorials which present a sign inversion of the T-LPE in the
off state followed by a nearly exponential relaxation back to equilibrium.
We have corroborated this behavior of the T-LPE response with numerical
simulations. We have also qualitatively explained the results with a simple
model which takes into account the local inductance of the metallic line
structure deposited on top of a Schottky barrier. We observe a substantial
increase of the position sensitivity of patterned line structures when their
width is reduced, measured by the peak to peak response.

The optical setup depicted in Fig. 1(a) comprises a microscope objective
lens (MO) (50x, 0.42 NA, Plan APO, working distance 21 mm) that focuses the
laser beam into the sample. The image of the sample is relayed into a CCD
camera by using the objective and a beam splitter. A diagram of the LPE
device examined in this study is also shown. The potential difference
created along the line is measured from three pairs of $500\times 500$ $\mu
m^{2}$ Cobalt pads which are contacted with gold wires by using indium. The
transient LPE has been studied by applying a train of periodic laser beam
pulses. TOPTICA-iBeam Smart diode lasers which emit light of 405 or 487 nm
of wavelength $\lambda$ have been used. The typical response of our device
to a laser pulse with $\lambda=487 nm$ is shown in Fig. 1(b), and is
discussed in detail below. Both wavelengths are compared in the supplemental
material \cite{SuppMat}.

The T-LPE signal between contacts was amplified in two stages. The first
stage is a home-made low noise preamplifier (bandwidth from DC to 1 MHz)
with a maximum gain of 47 is followed by a low noise SR750 amplifier with a
tunable gain and filter bandwidth. Fig. 1(c) shows that depending on the
roll-off frequency of the amplifier filter, the detection of the fast
``laser off'' transition can be affected. The voltage, amplified up to $%
10^{5}$ times, was measured using a NI-PCI 5922 digitizer which works at
frequencies up to 2MHz. The square wave used to modulate the laser pulses
was supplied by a Keithley K6221 current source. A number of experimental
parameters were controlled by computer software, such as the sample motion
(via Zaber T-L-A linear actuator), the frequency of the pulses, the power of
the laser beam and data acquisition.

A finite element analysis software ATLAS (by SILVACO) along with the LUMINUS
optoelectronic module has been used to simulate the LPE on multilayered
structures.

\begin{figure}[tbp]
\begin{center}
\includegraphics[width=8.5cm]
{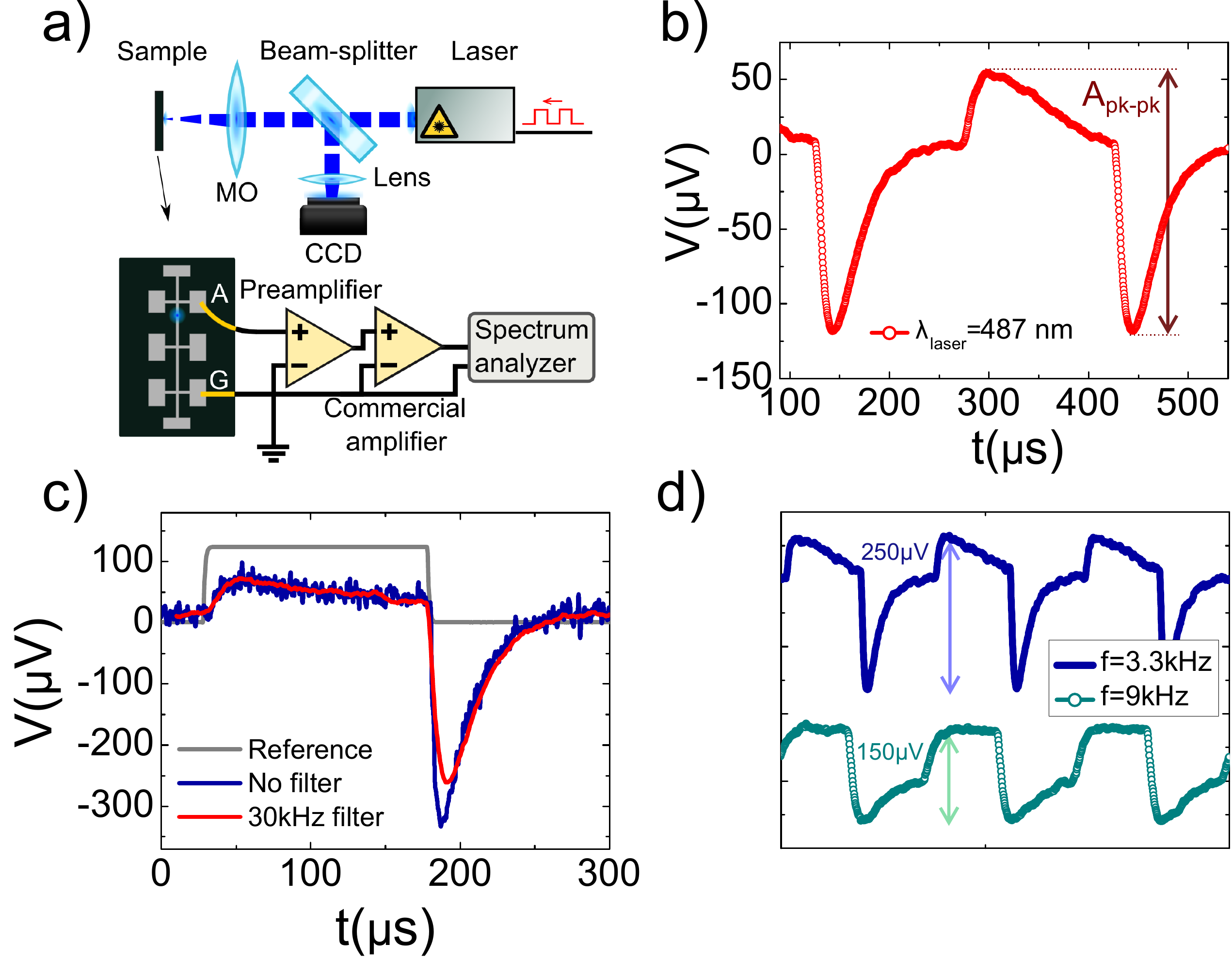} \label{Fig1}
\end{center}
\caption{(a) Diagram of the experimental set-up. (b) Time dependent LPE with
f=3.33 kHz and P=0.76mW. (c) Power dependence of the peak-to-peak values of
LPE taken with a pulse frequency of 3.33 kHz with the focused laser located
on top of the left contact. (d) Change of the response at a fixed laser
power with the frequency of the pulse (3.33 and 9.9 kHz) with laser spot
centered at 200 $\protect\mu m$ from the top electrode.}
\end{figure}

% %\section{Experimental results}

Figure 1(d) presents the T-LPE measured for a 10 $\mu m$ wide Co line at
different chopping frequencies. One notes that in the ON stage, the T-LPE
voltage first increases and then gradually decreases. In the OFF stage the
T-LPE voltage shows a fast sign inversion followed by slow, nearly
exponential decay down to zero. By using a low-pass filter we obtain a
better signal-to-noise ratio, which does not alter the measurement except
for the amplitude of the negative peak (see Fig. 1(c)). We found that the
optimum chopping frequency for which the T-LPE response is effectively
analyzed ranges around a few kHz. As shown in Fig. 1(d), the peak to peak ($%
pk-pk$) amplitude somewhat depends on the chopping frequency. This optimum
frequency range is restricted from below by the maximum digitizer rate and
from above by typical relaxation rate in T-LPE. Most of the experimental
results presented below correspond to T-LPE in 10 $\mu m$ wide Co line
measured with a 3.3 kHz chopping frequency and a focused beam spot size of 2 
$\mu m$. Measurements in 20 and 5$\mu m$ wide Co line structures show a
qualitatively similar T-LPE response, but with different $pk-pk$ amplitudes
(see below).

\begin{figure}[tbp]
\begin{center}
\includegraphics[width=8.5cm]
{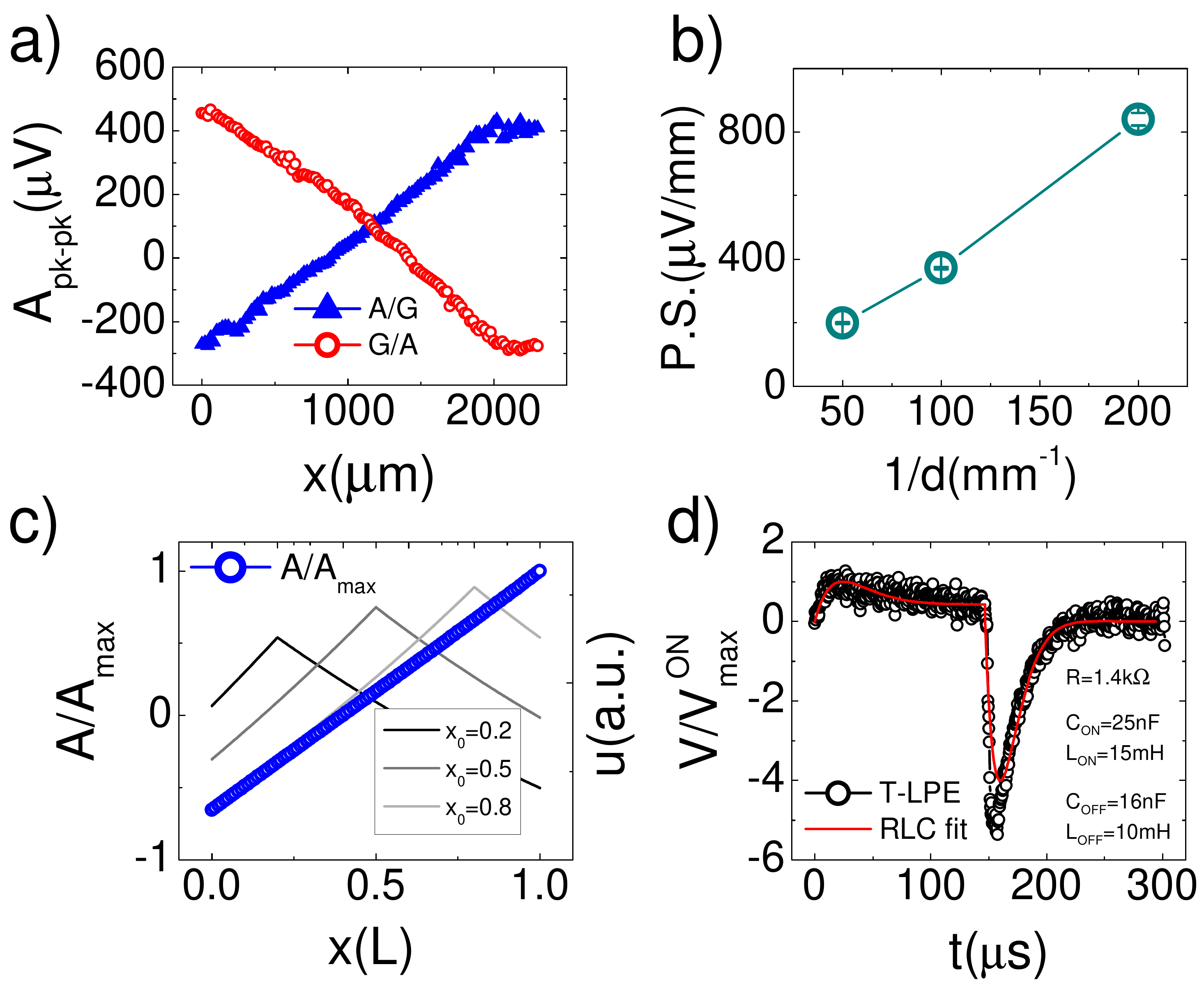} \label{Fig2}
\end{center}
\caption{(a) Dependence of the signal amplitude with the position X, for two
electrode configurations, showing a typical LPE linear behaviour. (b)
Dependence of the resistance and position sensitivity with $1/d$, with $d$
the Co line width. (c) Position-dependent solution of the model, where $x_0$
indicates where the laser spot is located. The constants used are: $%
B=1,F=1,E=1,D=3,L=1,h=0.6,s=-1$. (d) Comparison of the T-LPE response of a
10 $\protect\mu m$ Co line to the driven RLC circuit equation.}
\end{figure}

% \subsection{Qualitative model}

Figure 2(a) presents the $pk-pk$ amplitude of the T-LPE voltage as a
function of the position of the laser spot on top of the Co line, relative
to the distance from the reference (grounded) contact. The laser beam power
is P=2 mW, which corresponds to the end of the linear and the beginning of
the saturated response regime \cite{SuppMat}. Figure 2(b) shows the
estimated position sensitivity (P.S.) in $\mu V/mm$ for our 20 $\mu m$, 10$%
\mu m$ and 5 $\mu m$ wide samples, as a function of $1/d$ with $d$ the width
of the Co line. The sensibility of the samples is estimated from the slope
of a linear fit of the $pk-pk$ amplitude vs. position graphs (see for
example Figure 2(a)). More information on the analysis of the $pk-pk$
amplitude as a function of laser power can be found in the supplemental
material \cite{SuppMat}.

As can be seen in Fig. 2(b), the position sensitivity substantially improves
as the samples are made narrower. This suggests that making the PSD narrower
and using the peak to peak T-LPE response in such structured sensors could
be a simple and effective way to substantially improve their time-space
sensitivity. Among the factors which could limit the applicability of
microstructuring by optimizing the width are the laser spot dimensions and the characteristics
of the material. They could include a possible reduction of the diffusion of photocarriers or the
vulnerability of these narrow line structures to ambient conditions (e.g. edge
oxidation).

We observe that far from the electrodes that the LPE amplitude has a close
to linear dependence on the spot position (Fig. 2(a)), which is one of the
most important characteristics of previously studied planar LPE devices \cite%
{WALLMARK1957,Lucovsky1960,Boeringer1994,Zhao2006}. This behavior is
independent of the contact configuration with respect to the ground (Fig.
2(a)). Interestingly, however, the T-LPE amplitude and its peak to peak
values appear to show finite (above experimental resolution) values when
spot is centered exactly between the electrodes. We attribute this offset to
a difference in the boundary conditions for the diffusive carriers close the
electrodes. The grounded electrode (G) is described as having a small Co
contact resistance and the electrode that is connected to the preamplifier
(A) which has a much larger input resistance. We have been able to reproduce
qualitatively the existence of such an offset by using a one dimensional
drift-diffusion model \cite{WALLMARK1957,Lucovsky1960} which calculates the
stationary potential distribution (lines) and potential difference (dots)
employing two different (Robin / 3-rd type) boundary conditions (see Figure
2(c) and the supplemental material\cite{SuppMat}).

\begin{figure}[tbp]
\begin{center}
\includegraphics[width=8.5cm]
{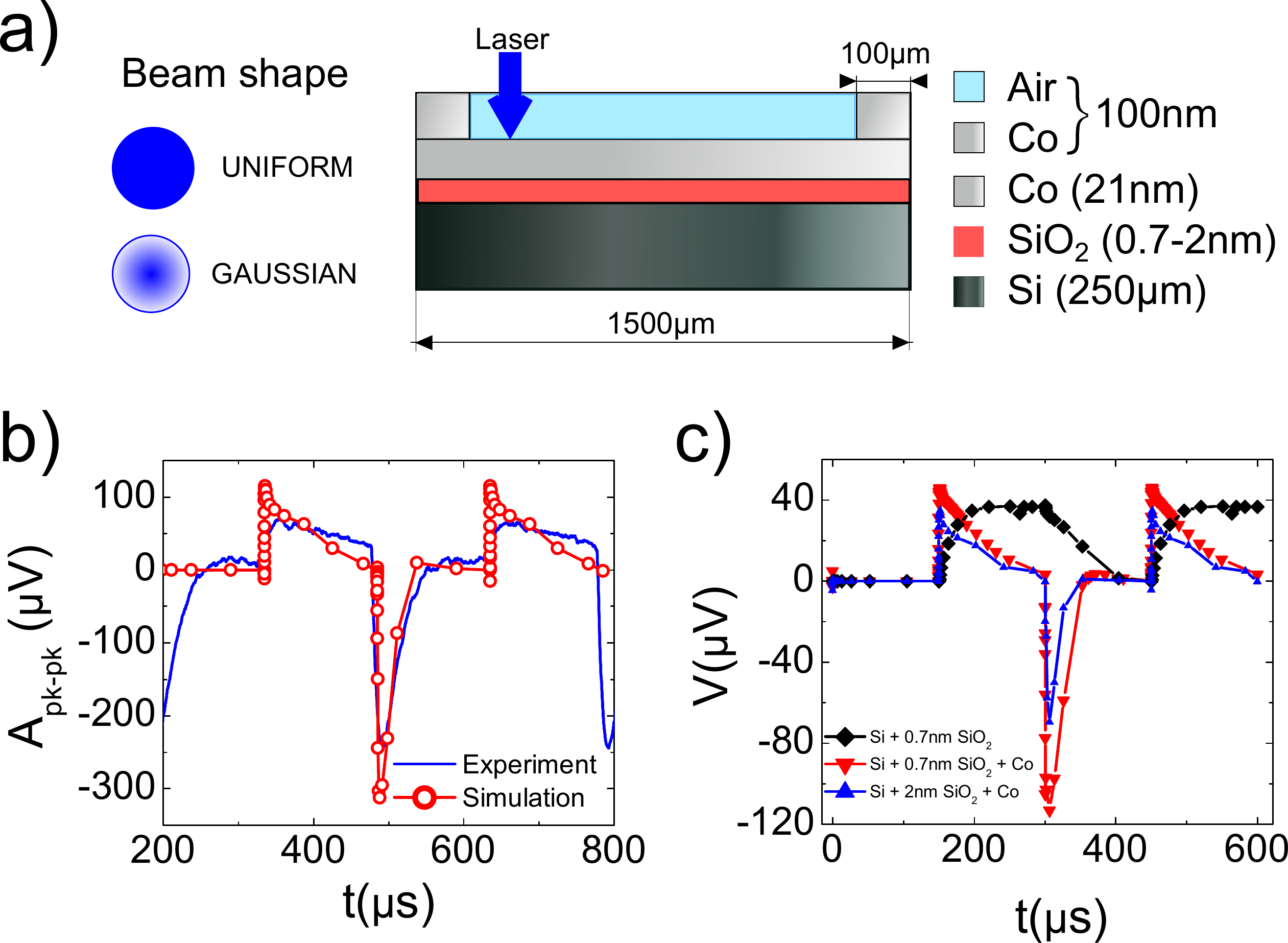} \label{Fig3}
\end{center}
\caption{(a) Sketch of the simulated layered structure with the different
beam shapes. (b) Comparison between the experiment and simulations of a Co
layer on SiO2-Si with a uniform beam. The parameters used are: concentration
of doping carriers in Si; n-type = $10^{13}cm^{-3}$ and p-type = $2\times
10^{13}cm^{-3}$; carrier's lifetimes $\protect\tau _{p}=2\times 10^{-5}$ $s$; 
$\protect\tau _{n}=10^{-6}$ $s$; Si electrical resistivity $\protect\rho =16$
$\protect\mu \Omega \times cm$. A Gaussian laser source was chosen with an
irradiance $I=1600$ $W/cm^{2}$, $\protect\lambda =405$ $nm$ and a modulating
frequency of $f=3.33$ $kHz$. (c) Dependence of the signal with the layer
structure. The experiment is reproduced only with Co in the structure. A
gaussian beam is used in these simulations.}
\end{figure}

An earlier model, which describes the dynamic T-LPE response in wide
two-dimensional structures \cite{WALLMARK1957,Lucovsky1960}, is capable only
of explaining the charge-discharge dynamic response similar to an RC circuit
(local resistance-capacitance). In order to explain the peak-like T-LPE and
its sign inversion in the off state we propose introducing an additional
term (second time derivative) in the differential equation for the potential
distribution $u(x,t)$. This new term corresponds to the local inductance $L$
due to presence of the metallic wire:

\begin{equation}
A\frac{d^{2}u(x,t)}{dt^{2}}+B\frac{du(x,t)}{dt}-D\frac{d^{2}u(x,t)}{dx^{2}}%
+Eu(x,t)=F(x,t)  \label{xtdepeq}
\end{equation}

where $A\propto L,C$, $B$ $\propto R,C$, the relation between terms $D$ and $%
E$ describes the carrier diffusion along the device and F corresponds to the
electron-hole separation function. The time dependent solution of equation
(1), at a fixed $x_0$, corresponds to a driven harmonic oscillator, or RLC
circuit:

\begin{equation}
L\frac{d^{2}u(x_{0},t)}{dt^{2}}+R\frac{du(x_{0},t)}{dt}+\frac{1}{C}
u(x_{0},t)=F(x_{0},t)
\end{equation}

We have found an excellent agreement with our experimental results, as shown
in Figure 2(d), for the case when equation 1 describes an underdamped
oscillator very close to the overdamped regime\cite{SuppMat}. Using the
resistance of the 10 $\mu m$ Co line, we obtain an estimate of the
capacitance and inductance of the strip from the fit (see Fig. 2(d)). The
resistance of the 5 $\mu m$ sample is twice that of the 10 $\mu m$ strip,
and using as capacitance half of what was obtained for the 10 $\mu m$
sample, we obtain values of inductance which are 60 \% higher\cite{SuppMat}.

Our numerical simulations, with adjusted parameters and a wavelength $%
\lambda =405nm$ have reproduced qualitatively the main experimental findings
as can be seen in Figure 3(a,b). The main simulation parameters which
resulted in the best fit to the experiment are shown below and discussed in
the supplemental material \cite{SuppMat}. We attribute some differences
between the experiment and simulation to the capacitance of a real sample as
well as the influence of the pre-amplifier, absent in the simulations. In
order to check if the observed effects are specific to MOS structures, we
have also simulated the T-LPE under the same conditions by decreasing the SiO%
$_{2}$ layer thickness (or removing it altogether), shown in Fig. 3(c). When
we remove both SiO$_{2}$ and Co layers and leave only the laterally
contacted Si surface, the transient LPE response of the pure Si substrate
shows an exponential increase (decrease) when the laser is switched ON (OFF)
(Fig. 3(c)).

In conclusion, we have studied the transient lateral photoeffect in
patterned metal-oxide/semiconductor structures with widths of several
microns. The dependence of the T-LPE with the spot position is almost linear
and the time dependent response shows a sign inversion after the laser is
switched off. The dynamic response has been explained by the influence of a
local inductance, in addition to what was previously considered for wider
LPE devices: a local capacitance (generation/recombination process) and
resistance. Our findings could stimulate the development of micron-wide
position sensitive detectors with improved time-space resolution needed in
microrobotics \cite{Ivan2012} and other fields where a precise control of
the trajectory is required.

The authors acknowledge A. Levanuyk for his interest and fruitful
discussions, L. Martin for her simulations on the initial stages of the work
and Ch. van Haesendock for preparing the samples. This work has been supported by the 
Spanish MINECO (MAT2012-32743) and Comunidad de Madrid (P2009/MAT-1726) grants.\\

* Corresponding author: farkhad.aliev@uam.es

\end{document}